# Spin Splitting Induced by a Competition between Quantum Spin Hall Edge States and Valley Edge States

Mudan Yang, Jia-Bin Qiao, Zhao-Dong Chu, Wen-Yu He, Lin He*

Department of Physics, Beijing Normal University, Beijing, 100875, People's Republic of China

**Strained graphene with lattice deformations has been demonstrated to give rise to large pseudomagnetic fields and host many exotic properties. Here, we propose a non-magnetic approach to realize a momentum-dependent out-of-plane spin splitting in strained graphene nanoribbons with a moderate spin-orbit coupling. This unique spin splitting distincts from the well-known Zeeman-type spin splitting and the Rashba-type spin splitting. Our analysis indicates that the competition between quantum spin Hall edge states and valley edge states in the nanoribbon leads to the unique spin splitting. The quantum spin Hall states at one edge of the nanoribbon are suppressed by the counterpropagating edge modes ($K$ and $K'$ valleys) induced by the pseudomagnetic field. At the opposite edge, the quantum spin Hall states are not affected at all. Therefore, the degenerate quantum spin Hall states of opposite spin orientation, which propagate at the two opposite edges of the nanoribbon, are lifted. This result reveals a new method to manipulate the spin degrees of freedom of electrons.**

Usually, we have to apply an external magnetic field or use magnetic dopants, which break the time-reversal symmetry, to generate Zeeman-type out-of-plane spin splitting. The ability to tailor the spin splitting in the absence of magnetic field or any explicit time-reversal-invariant breaking effect motivates a sizable fraction of modern research in condensed matter physics [1-5]. However, a non-magnetic approach to realize the Zeeman-type spin polarization has proved to be a challenging task because of that it violates Kramers theorem. According to the Kramers theorem, two electron states with opposite momenta and spin must form a Kramers doublet and consequently the Zeeman-type spin splitting is forbidden in a time-reversal-invariant system.

In graphene and other two-dimensional atomic crystals with honeycomb lattices, the above issue can be circumvented, i.e., we can satisfy the Kramers theorem and generate Zeeman-like spin splitting simultaneously without introducing any time-reversal-invariant breaking effect [6-8]. The honeycomb lattices of the two-dimensional atomic crystals result in two independent Dirac cones, commonly called K and K′ valley, centered at the opposite corners of the hexagonal Brillouin zone [9]. Because of K′ = -K, electron states in valley K are transformed into states in valley K′ under time reversal. Therefore, it is possible to generate the Zeeman-like spin splitting without violating the Kramers theorem when the spin polarizations are in opposite directions in the two valleys. Very recently, a possible route to realize such a unique spin splitting has been proposed in deformed graphene nanoribbons with a moderate spin-orbit coupling (SOC) [6], and, almost simultaneously, the unique Zeeman-like spin splitting has been demonstrated experimentally in WSe$_2$ sheets [7]. In monolayer WSe$_2$, the charge on the W atoms and Se atoms generates local in-plane dipolar interaction causing a strong SOC, which mimics an out-of-plane magnetic field and induces the momentum-independent out-of-plane spin splitting around K and K′ valleys (the generated spin polarization around K valley is opposite to that around K′ valley to preserve the time-reversal symmetry.) [7,8]. Such an effect is expected to be transferred to other graphene sheet analogues without inversion symmetry, for example monolayer $MoS_2$.

In strained graphene nanoribbons, the obtained spin splitting is out-of-plane but momentum-dependent [6], which differs from the Zeeman-like spin splitting observed in $WSe_2$ sheets and also distincts from the Rashba-type spin splitting with a momentum-dependent in-plane spin polarization [7,8]. It was proposed that the strain-induced pseudomagnetic field in graphene lifts the degenerate quantum spin Hall (QSH) edge states of opposite spin orientation. Because of opposite signs of the pseudomagnetic field in two valleys, the resulting spin polarizations are in opposite directions in the K and K′ valleys of graphene [6]. In this paper, we further study the unique spin splitting in strained graphene nanoribbons. According to our analysis, the pseudomagnetic field seems to selectively affect the QSH edge states. At one edge of the nanoribbon, the QSH states are suppressed by the counterpropagating edge modes ($K$ and $K′$ valleys) induced by the pseudomagnetic field. At the opposite edge of the nanoribbon, the QSH states are not affected at all. Therefore, the degenerate QSH states of opposite spin orientation, which propagate at the two opposite edges of the nanoribbon, are lifted. This result provides a new method to generate spin polarization in two-dimensional honeycomb structures through spin-valley coupling.

The tight-binding model Hamiltonian of graphene with the SOC can be written as [10,11]

$$H = \sum_{\langle ij \rangle \alpha} t c^{\dagger}_{i\alpha} c_{j\alpha} + \sum_{\langle\langle ij \rangle\rangle \alpha\beta} i t_2 v_{ij} s^z_{\alpha\beta} c^{\dagger}_{i\alpha} c_{j\beta}. \qquad (1)$$

Here, the first term is the nearest neighbor hopping term on the honeycomb lattice, $t$ is the hopping integral, and the operators $c^{\dagger}_{i\alpha}$ ($c_{i\alpha}$) create (annihilate) an electron with spin $\alpha$ at site $i$. Surprisingly, this extremely simple model of the first term is able to describe correctly the electronic structure of graphene in most situations [9]. Around the Dirac points, the low-energy excitations of graphene have a linear dispersion and the first term of Hamiltonian (1) becomes two-dimensional massless Dirac equation. In strained graphene, lattice deformations change the electron hopping between sublattices and some kinds of hopping modulations gives rise to an effective gauge field **A** in the low-energy Dirac equation [12,13]. The gauge field has been conceptually predicted to affect in-plane motion of Dirac fermions and mimic out-of-plane pseudomagnetic fields $B_S$ as large as 100 T. Such an effect has been experimentally demonstrated in strained graphene by local scanning tunneling microscope technique [14-19], and many interesting properties are further proposed to be realized in the strained graphene [20-28]. Figure 1(a) shows a concrete hopping modulation to realize a uniform pseudomagnetic field with $A_x = \frac{c}{ev_F}(t - t_1) = B_S y$ and $A_y = 0$ [6,28]. The pseudomagnetic field results in flat bands in the electronic structure of graphene at discrete energies, which are similar to the Landau levels generated in real magnetic fields [12-19]. However, the strain-induced gauge field and the pseudomagnetic field have opposite signs in the valleys K and K′ to preserve the time-reversal symmetry.

The second term of Hamiltonian (1) describes the spin-orbit interaction, which is an

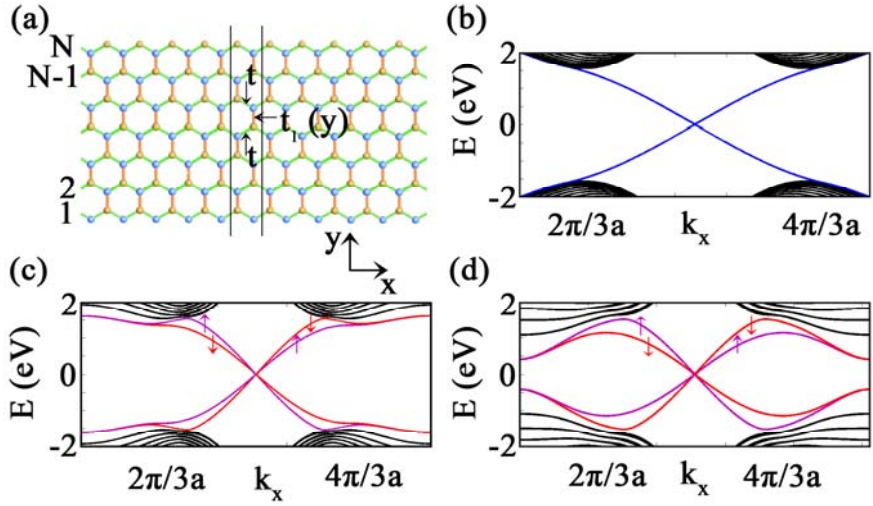

**Figure 1** (color online). (a) Diagram of a zigzag graphene nanoribbon. *N* is the number of zigzag chains in the nanoribbon. The hopping matrix element on orange bonds along the *y* direction is changed as $(t_1-t) = ev_F B_S y/c$ to induce a uniform pseudomagnetic fields. (b) One-dimensional band structure of a zigzag graphene nanoribbon with *N* = 60 and $t_2/t = 0.1$. The degenerate edge states, in which up and down spins propagate in opposite direction, connect the valence band to the conduction band. (c) A pseudomagnetic field $|B_S|$ = 15.2 T lifts the degenerate quantum spin Hall edge states of opposite spin orientation. Purple and red curves represent spin-up and spin-down. The spin splitting is out-of-plane and momentum-dependent. (d) A larger pseudomagnetic field $|B_S|$ = 24.3 T further lifts the degenerate quantum spin Hall edge states of opposite spin orientation.

effect of relativistic origin that couples electron spin and orbital momentum [10,11]. Here, $t_2$ is a spin dependent second neighbor hopping, $v_{ij} = \pm 1$ depending on the orientation of the two nearest neighbor bonds the electron traverses in going from site $j$ to $i$, and $s^z_{\alpha\beta}$ is the Pauli matrix describing the electron's spin [10,11]. Figure 1(b) shows the QSH states for a non-strained zigzag graphene nanoribbon (i.e., $t_1 = t$) obtained by solving Hamiltonian (1). In the QSH phase, edge states with opposite spin counterpropagate at the edges of graphene nanoribbon. Additionally, the energies of the clockwise and counterclockwise edge channels are degenerate, as shown in Fig. 1(b). However, a pseudomagnetic field can lift the degenerate QSH edge states [6]. The pseudomagnetic field, which is a unique out-of-plane orbital field [6,29], affects the clockwise and counterclockwise edge currents of the QSH states, where the direction of motion was determined by the spin orientation. Consequently, it lifts the degenerate edge states of opposite spin orientation. The coexistence of the valley-dependent pseudomagnetic fields and the SOC results in a unique out-of-plane momentum-dependent spin polarization in graphene nanoribbons, as shown in Fig. 1(c) and Fig. 1(d) [30].

Here, we should point out that the pseudomagnetic field seems to selectively affect the QSH edge states in graphene nanoribbons. For $E > 0$, the pseudomagnetic fields only lower (raise) the energies of edge states with spin down (spin up) for the states with $k_x < \pi/a$ ($k_x > \pi/a$), i.e., in the $K$ ($K'$) valley. A larger pseudomagnetic field can further reduce the group velocity of edge states with spin down for the states with $k_x < \pi/a$, as shown in Fig. 1(d). However, the edge states with spin up for the states with $k_x > \pi/a$ is not affected by the pseudomagnetic field equivalently, as shown from Fig. 1(b) to Fig. 1(d). This asymmetric

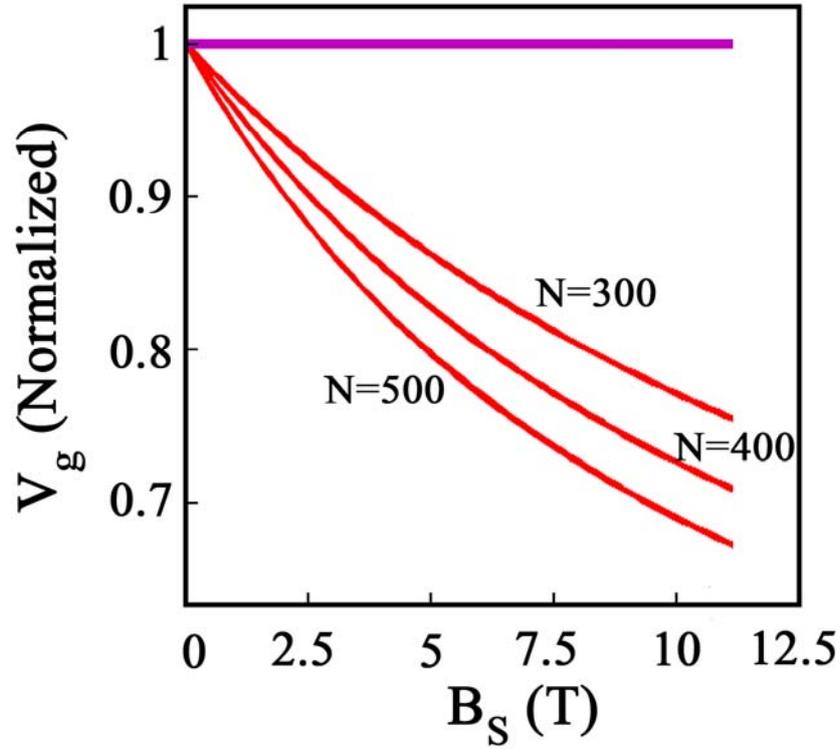

**Figure 2** (color online). The reduced group velocities $V_g$ of the QSH edge states with $E > 0$ and $k_x < \pi/a$ as a function of pseudomagnetic field. Purple and red curves represent spin-up and spin-down edge states, which propogate at opposite edges of the graphene nanoribbons. The group velocities of the spin down edge states decrease with increasing the pseudomagnetic field and the group velocities decreases quicker for graphene nanoribbons with larger $N$. However, the group velocities of the spin up edge states is almost not affected by the pseudomagnetic field at all.

spin

spin splitting between the spin up and spin down states distincts from the symmetric Zeeman spin splitting induced by a real magnetic field.

To further illustrate the asymmetric spin splitting observed in strained graphene nanoribbons, we plot the group velocities $V_g$ at $E = 0^+$ for the two QSH edge states with $E > 0$ and $k_x < \pi/a$ as a function of pseudomagnetic field, as shown in Fig. 2. The two states with opposite spin orientation propogate at different edges of the graphene nanoribbon. With increasing the pseudomagnetic fields, the group velocities for the spin down edge states decrease. However, the group velocities for the spin up edge states almost keep a constant, which means that this QSH edge state is not affected by the pseudomagnetic fields at all. Similar result can be obtained for edge states with $k_x > \pi/a$ and $E > 0$ if we switch the spin orientations of the occupying electrons. In the QSH states, electrons, which have both the opposite spin orientation and the opposite propagating directions, are supported by the same edge of the nanoribbon [10,11]. Therefore, we can conclude that only the QSH states at one edge of the nanoribbon are suppressed by the pseudomagnetic fields, the QSH states at the opposite edge are not affected at all.

The above result indicates that the pseudomagnetic field destroys the spatial symmetry of the opposite edges of the graphene nanoribbon and consequently breaks the symmetry of the QSH states propagating at the two edges. The reason why the effect of pseudomagnetic field is absent at one edge whereas it is very obvious at the opposite edge of the nanaoribbon can be explained by the competition between the QSH effect and the valley edge states induced by the pseudomagnetic fields. Figure 3 shows schematics of the competition between the edge states of QSH and the valley edge states. Because of the

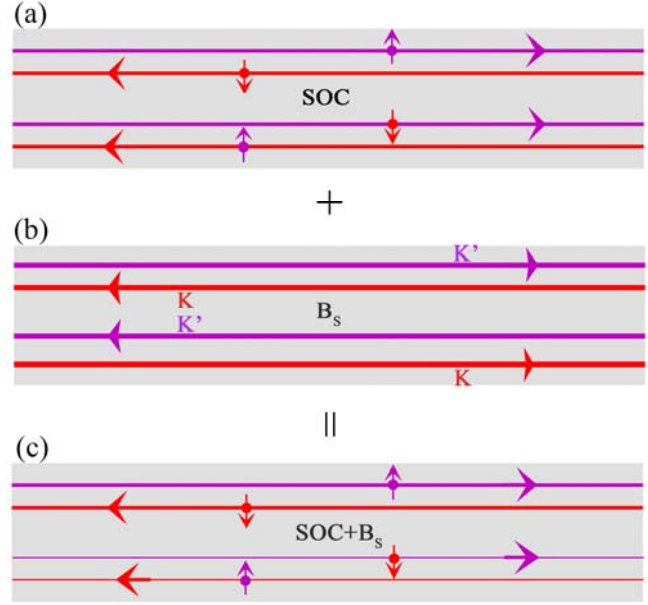

**Figure 3** (color online). (a) Sketch of the edge currents in QSH state of a ziazag graphene nanoribbon. The upper edge supports a forward mover with spin up and a backward mover with spin down and conversely for the lower edge. Purple and red lines represent edge states in the $K'$ valley (with $k_x > \pi/a$) and the $K$ valley (with $k_x < \pi/a$), respectively. (b) Sketch of the edge currents of a ziazag graphene nanoribbon induced by a pseudomagnetic field. The pseudomagnetic field generates counterpropagating edge states, where each direction is associated to a given electron valley. (c) Competition between the QSH edge states and the valley edge states in the nanoribbon. In the upper edge, the edge currents of the QSH and the valley edge states from the same valley (either $K$ or $K'$ valley) are propagating in the same direction. Whereas, in the lower edge, they are propagating in the opposite direction. Therefore, the QSH edge states in the lower edge are partially suppressed by the edge currents induced by the pseudomagnetic field.

opposite signs in valleys *K* and *K'*, the pseudomagnetic field generates counterpropagating edge states, where each direction is associated to a valley, as sketched in Fig. 3(b). In the QSH state of a zigzag graphene nanorbibbon, the edge states in a fixed energy, which have opposite spin orientation and propagate in the same direction, are from the same valley, as shown in Fig. 3(a). When both the QSH effect and the valley edge states coexist, the edge currents of the QSH and the valley modes from the same valley (either *K* or *K'* valley) are propagating in the opposite direction in one edge of the nanoribbon, as shown in Fig. 3(c). Then, the QSH edge states at this edge is suppressed by the valley edge modes generated by the pseudomagnetic field. Simultaneously, the edge states at the opposite edge is not affected at all, as shown in Fig. 3(c). As a consequence, the degenerate QSH states of opposite spin orientation, which propagate at the two opposite edges of the graphene nanoribbon, are lifted.

According to the result shown in Fig. 2, the group velocities of the spin down edge states almost decrease linearly with the pseudomagnetic field. For small pseudomagnetic fields, the dependency of $V_g$ and $B_S$ can be approximated as $V_g = 1 - P \times B_S$. It is interesting to find that the slope *P* does not depend on the magnitude of SOC in the graphene nanoribbons (not shown). It only depends on the width of the nanoribbon, i.e., the number of *N* in our model, as shown in Fig. 2. Figure 4 shows the slope as a function of *N*. Obviously, the value of |*P*| increases linear with *N*. Both the independence of the SOC and the linear dependence of *N* for the slope |*P*| could be well understood according to the mechanism proposed in Fig. 3. In the QSH states, a two-terminal measurement on the graphene nanoribbon would give the Hall conductance $\sigma_{xy} = \dfrac{2e^2}{h}$ [10,11,31-34]. The

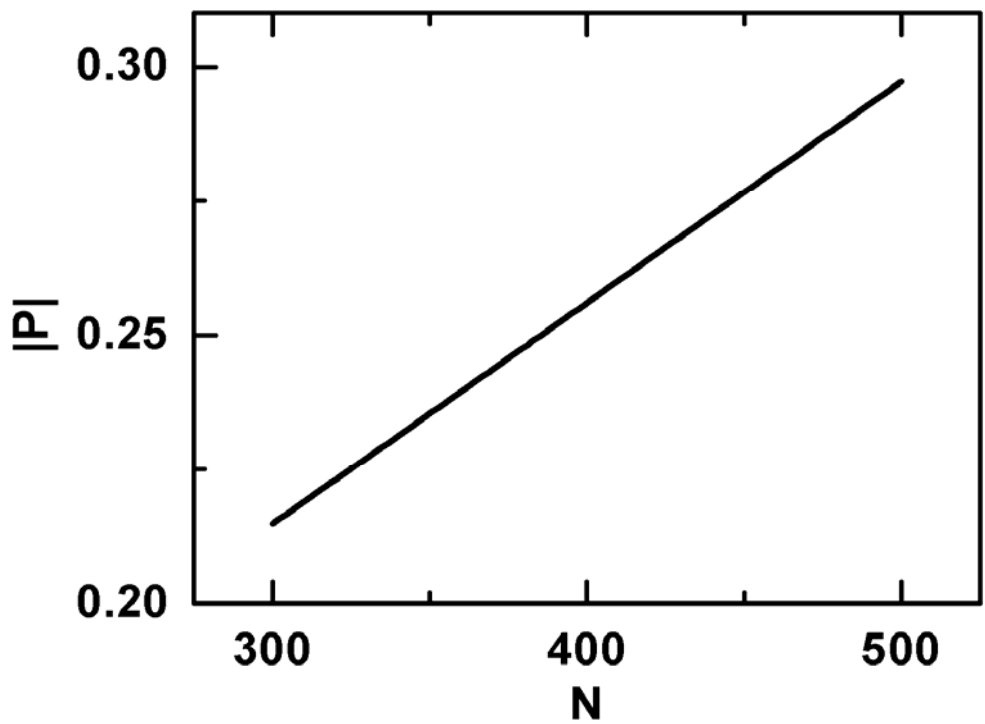

**Figure4** (color online). The slope $P$ as a function of $N$. $N$ is the number of zigzag chains in the nanoribbon.

conductance of each QSH edge channel $\sigma_{xy}^{QSH}$ should approach a value close to $\frac{e^2}{h}$ and is independent of the magnitude of SOC. In strained graphene, the Hall conductivity induced by the pseudomagnetic field for each valley is given by [35,36]

$$\sigma_{xy}^{valley} = -\frac{1}{2}\frac{e^2 c\,|\rho|\,\mathrm{sgn}(eB_S)\mathrm{sgn}\,\mu}{|eB_S|} \sim \frac{1}{SB_S}. \qquad (2)$$

Here $\rho$ is the two-dimensional electronic density, $\mu$ is the chemical potential, $S$ is the area of the nanoribbon, 1/2 arises from the degeneracy of the $K$ and $K'$ valleys. The QSH edge states at one edge of the nanoribbon are completely suppressed when $\frac{1}{2}\sigma_{xy}^{valley} = -\sigma_{xy}^{QSH} \equiv -\frac{e^2}{h}$. Here, the minus means that the two edge currents at one edge of the nanoribbon are in opposite directions. We define a critical pseudomagnetic field $B_S^C$, at which the group velocities for one of the QSH edge states is reduced to exactly zero. Therefore, the critical pseudomagnetic field $B_S^C \sim \frac{1}{S} \sim \frac{1}{N}$. According to the relation $V_g$ = 1-$P \times B_S$ obtained in Fig. 2, we can obtain $|P| \propto \frac{1}{B_S^C}$. Consequently, we have $|P| \propto N$ and the magnitude of $P$ is independent of SOC in the graphene nanoribbon.

The spin-valley coupling in zigzag graphene nanoribbons, as revealed in this paper, results in an asymmetric quantum spin Hall (AQSH) state: the QSH states propagating at the two edges is asymmetry. In the AQSH state, the spin-polarized edge states propagate without dissipation at one edge of the nanoribbon and the edge currents at the opposite edge are partially suppressed. The pre-existing quantum Hall related effects include the quantum

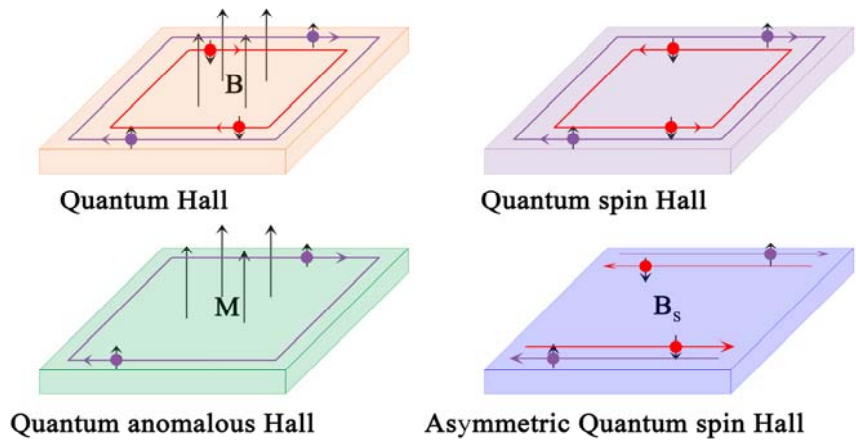

**Figure 5** (color online). Quantum Hall quartet. Schematic of the edge channels in quantum Hall related effects. The locking schemes between spin and flow direction of electrons, and the number of edge channels depend on the material details. Here only the simplest cases are illustrated here.

Hall effect, the QSH effect, and the quantum anomalous Hall effect [10,11, 31-38]. The schematic figure, as shown in Fig. 5, shows the complete quantum Hall quartet and the schematic of the edge channels of the quantum Hall related effects. Our analysis points out that the pseudomagnetic field breaks the spatial symmetry of the QSH states propagating at the two edges and leads to the AQSH state. This result provides a unique method to tune the QSH edge currents in two-dimensional atomic crystals with honeycomb lattices through the strain engineering.

In summary, we report a unique zero-field spin splitting induced by the competition between QSH edge states and the valley edge states. Our result indicates that the valley edge states generated by the pseudomagnetic field can tune the QSH edge states at one edge of the graphene nanoribbon and consequently breaks the spatial symmetry of the QSH states propagating at the two edges. This result not only provides a non-magnetic approach to generateunique spin polarization in graphene and other graphene sheet analogues but also adds a new type of quantum Hall related effects.

**Acknowledgements**

The authors would like to thank Prof. Jian Wang, and Prof. Yanxia Xin for helpful discussions and comments. We are grateful to National Science Foundation (Grant No. 11374035, No. 11004010), National Key Basic Research Program of China (Grant No. 2013CBA01603, No. 2014CB920903), and the Fundamental Research Funds for the Central Universities. Mudan Yang, Jia-Bin Qiao, and Zhao-Dong Chu contributted equally to this paper.

*Email:helin@bnu.edu.cn.